\documentclass[aps,prl,superscriptaddress,showpacs,twocolumn]{revtex4}
\usepackage{graphicx}
\usepackage{dcolumn}
\usepackage{bm}
\usepackage{graphicx}
\usepackage{amsmath, amsfonts, amssymb,mathrsfs}
\usepackage{pstricks}
\usepackage{amsxtra}
\usepackage{amsthm}
\usepackage{natbib}
\def\be{\begin{equation}}      
\def\ee{\end{equation}}
\def\beu{\begin{equation*}}   
\def\eeu{\end{equation*}}

\providecommand{\abs}[1]{\left\lvert#1\right\rvert}   
\providecommand{\ket}[1]{\left|#1\right\rangle}

\providecommand{\bra}[1]{\left\langle#1\right|}
\providecommand{\mean}[1]{\left\langle#1\right\rangle}
\providecommand{\comm}[2]{\left[ #1, #2 \right]}  		

\graphicspath{{figures/}}
\begin{document}
\title{Nanoplasmonic  Lattices for Ultracold Atoms}
\author{M. Gullans}
\affiliation{Department of Physics, Harvard University, Cambridge, MA 
02138, USA}
\author{T. Tiecke}
\affiliation{Department of Physics, Harvard University, Cambridge, MA 
02138, USA}
\affiliation{MIT-Harvard Center for Ultracold Atoms, and Research Laboratory of Electronics, MIT, Cambridge, Massachusetts 02139, USA}
\author{D. E. Chang}
\affiliation{
ICFO-Institut de Ciencies Fotoniques, Mediterranean Technology Park, 08860 Castelldefels (Barcelona), Spain}
\author{J. Feist}
\affiliation{Department of Physics, Harvard University, Cambridge, MA 
02138, USA}
\author{J. D. Thompson}
\affiliation{Department of Physics, Harvard University, Cambridge, MA 
02138, USA}
\author{J. I. Cirac}
\affiliation{Max-Planck-Institut f\"ur Quantenoptik, Hans-Kopfermann-Str.~1, D-85748 Garching, Germany}
\author{P. Zoller}
\affiliation{Institute for Theoretical Physics, University of Innsbruck, 6020 Innsbruck, Austria}
\author{M. D. Lukin}
\affiliation{Department of Physics, Harvard University, Cambridge, MA 
02138, USA}
\date{\today}
\begin{abstract}
We propose to use sub-wavelength  confinement of light associated with the near field of plasmonic systems  to create nanoscale optical lattices for ultracold atoms. Our approach combines  the unique coherence properties of isolated atoms  with the sub-wavelength manipulation and strong light-matter interaction associated with nano-plasmonic systems.  It allows one to considerably increase the energy scales in the realization of Hubbard models and to engineer effective long-range interactions in coherent and dissipative many-body dynamics. Realistic 
imperfections and potential applications are discussed.   
\end{abstract}
\maketitle

Coherent optical fields provide a powerful tool for manipulating ultracold atoms \cite{BlochDalibard08,grimm99}.  However, diffraction sets a fundamental limit for the length-scale of such  manipulations, given by the wavelength of light \cite{hecht}. 
In particular, the large period of optical lattices determines the energy 
scale of the  associated  many-body atomic states \cite{buluta09,jaksch05,Leung12,lewenstein07}. The resulting scaling 
can be best understood by noting that  in the first Bloch band the maximum atomic momentum $\sim1/\ell$, where $\ell$ is the lattice spacing.  This sets the maximum kinetic energy to $~h^2/ m \ell^2$ \cite{Jaksch98}.  For conventional optical lattices the lattice spacing is set by half the wavelength of the trapping light $\sim 500$ nm; this yields corresponding tunneling rates of up to a few tens of kHz.   Additionally, for atoms in their electronic ground states interactions are restricted to short range.

Recent experimental \cite{Stehle11} and theoretical \cite{Murphy09,Chang09} work has demonstrated that integrating plasmonic systems with cold atoms represents a promising approach to achieving subwavelength control of atoms.  In particular, the experiments of Ref.~\cite{Stehle11} showed that ultracold atoms can be used to probe the near fields of plasmonic structures, paving the way to eventually trap atoms above such structures.
In this Letter we propose and analyze  a novel approach to the realization of high-density optical lattices using the  optical potential formed from the near field scattering of light by an array of plasmonic nanoparticles.  By bringing  atom trapping into the subwavelength and nanoscale regime we show that the intrinsic scales
of tunneling and onsite interaction for the Hubbard model can be increased by several  orders of magnitude compared to conventional optical lattices.  In addition, 
 subwavelength confinement of the atoms results in strong radiative interactions with the plasmonic modes of the nanoparticles \cite{deleon12}.   The coupled atom-plasmon system can be considered as a scalable cavity  array that results in strong, long range spin-spin interactions between the atoms with both dissipative and coherent contributions \cite{scalableCavityQED,Kimble08}.   
Such a system can be used for entanglement of remote atoms as well as  for novel realizations of coherent and  dissipative many-body systems.

\begin{figure*}[tbp]
\begin{center} 
\includegraphics[width = .99 \textwidth]{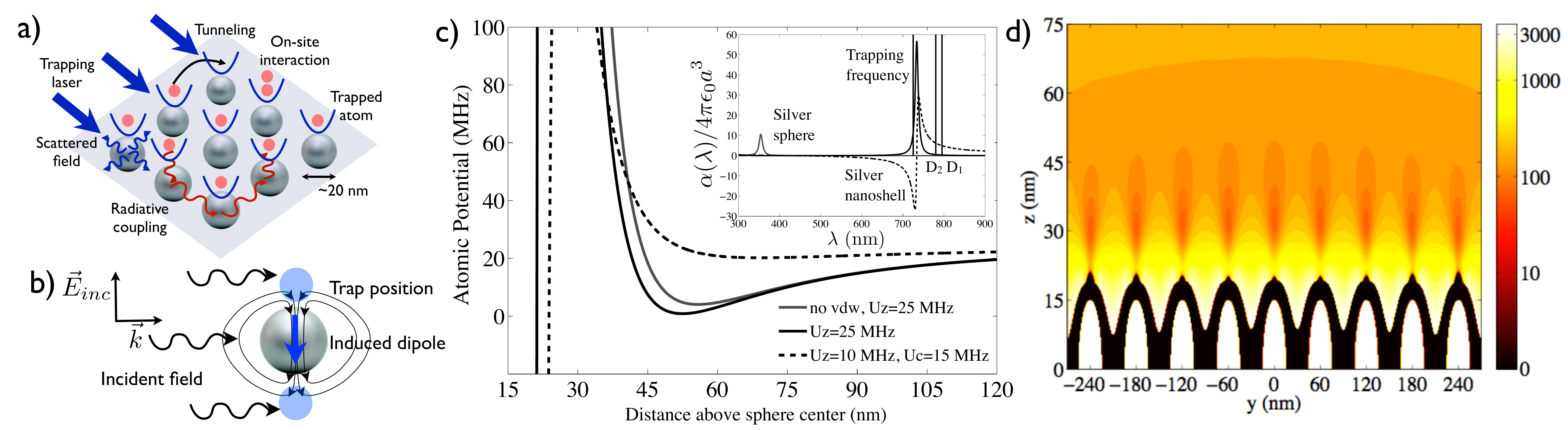}
\caption{\label{fig:fig1}
 a)  Illustration of the relevant physics in the plasmonic lattice. b)  Illustration of how to engineer a blue-detuned optical dipole trap by driving on the blue side of the plasmon resonance.  c)  Atomic potential for Rb including van der Waals (vdw) for trapping above a single silver nanoshell.  
 Dotted line shows how to weaken the trap by applying circularly polarized light perpendicular to the trapping light.  (Inset) Real (dashed) and imaginary (solid) part of the dipole polarizability  for a sphere and the nanoshell with a 15 nm radius and 13.85 nm SiO$_2$ core.  d) $y$-$z$ contours of atomic potential in MHz for a line of nine spheres in the center of a 45x45 square lattice with a 60 nm lattice spacing, black regions are where the potential is negative due to vdw, spheres are shown in white.  The nanoshells are silver with a 15 nm radius and 13.65 nm SiO$_2$ core, the trapping light is red detuned (704 nm) wrt to the plasmon resonance (682 nm) and applied from above with rotating $x$-$y$ polarized light.   
  }
\label{default}
\end{center}
\end{figure*}

To illustrate our approach  we first consider a single metallic nanosphere in vacuum illuminated by a plane wave.  For spheres small compared to a wavelength the dominant contribution to the scattered field is the dipole term, where the induced dipole moment  is given by $\bm{p} = \alpha(\omega) \bm{E}_0$ with
\begin{align}
\alpha(\omega)&=4 \pi \epsilon_0 a^3 \frac{\varepsilon(\omega)-1}{\varepsilon(\omega)+2}
\end{align}
where $a$ is the radius of the sphere and $\varepsilon$ is the permitivity \cite{Jackson}. 
The total electric field is
\be
\bm{E}= \bm{E}_0 + \frac{\alpha(\omega)}{4 \pi \epsilon_0} \frac{3 (\hat{r} \cdot \bm{E}_0) \hat{r} - \bm{E}_0}{ r^3} 
\ee
Near $\varepsilon(\omega_{sp})= -2$ there is a plasmon resonance and the scattered field can be engineered to create an optical dipole trap as depicted in Fig.~1(b).  
  Specifically, when the applied field is linearly polarized on the blue side of the plasmon resonance then
 the induced dipole will be $\sim\pi$ out phase with the incident field, leading to two intensity minima along the polarization direction at the positions $z_T^3 = \pm 2 a^3 \omega_{sp}^2/(\omega^2-\omega_{sp}^2)$, where we took a Lorentzian polarizability near the resonance $\alpha(\omega)=4 \pi \epsilon_0 a^3 \omega_{sp}^2/(\omega^2_{sp}-\omega^2 - i \omega \kappa)$, with $\kappa$ the linewidth.  For red detuned, circularly polarized light, there will be two minima along the propagation axis.
An atom can be trapped in these intensity minima via optical dipole forces 
\cite{grimm99}.  The trapping potential is given by $\hbar \Omega^2/\delta$, where $\Omega=\bm{\mu}_0 \cdot \bm{E}/\hbar$ is the Rabi frequency, $\bm{\mu}_0$ is the atomic dipole moment, and $\delta=\omega_a-\omega$ is the detuning between the atom and laser.  Expanding near the trap minima gives the trapping frequency $\omega_T^2 =  9 \frac{\hbar \Omega_0^2}{\delta \, m \, z_T^2} \textrm{Re}(\alpha)^2/\abs{\alpha}^2 \sim \hbar\, \Omega_0^2/ \delta\, m\, a^2$.

The trap depth can be controlled by applying a second field with the opposite polarization, 
as illustrated in Fig.~1(c).  Using this method, the atoms can be loaded into the near field traps by starting with a cold, dense gas of atoms in a large trap and then adiabatically turning on the near field traps.  

We now address several practical considerations. First, for alkali atoms there is a large disparity between the natural plasmon resonance and the atomic trapping transitions.  For a solid silver sphere the plasmon resonance occurs near 350 nm \cite{Johnson72Palik85}, compared to 780 nm for the D2 line in Rb.  However, the plasmon resonance is easily tuned by changing the geometry.   
Adding an inert core, such as SiO$_2$, will shift the plasmon resonance into the red \cite{Bohren83}, as illustrated in the inset to Fig.~1(c).  

There will also be significant surface interactions.  Van der Waals (vdw) forces can be overcome with modest laser power because of the sphere's plasmonic enhancement 
\cite{Murphy09,Chang09}.   
There are two dominant sources of heating and decoherence arising from incoherent transitions induced by the trapping laser or thermal magnetic field noise in the nanoparticle.  The first effect scales as $\gamma\, \Omega^2/\delta^2$, where $\gamma$ is the atomic linewidth, and is suppressed at large detuning.  
To estimate the effect of magnetic field noise we approximate the nanoshell as a current loop of radius and height $a$, thickness $t$, and resistivity $\rho$.  Then the incoherent transition rate between hyperfine states is $\sim (g_F \mu_0 \mu_B)^2 k_B T (a^4 t /r^5) /\hbar^2 \rho\, r  $, where $r$ is the distance of the atom to the sphere center, $g_F$ is the hyperfine g-factor, $\mu_B$ is the Bohr magneton, and $T$ is the temperature \cite{Henkel99}.   

Figures 1(c-d) show the atomic trapping potential for a single sphere and an array, respectively.  We numerically obtained the trapping potential in Fig.~1(c) using Mie theory and the vdw potential was obtained using the methods in Ref. \cite{Johnson11}.  To solve for the trapping potential in the array in Fig.~1(d) we approximated the scattered field from each nanoshell by a dipole and solved for the total field self-consistently.
  Using the parameters in Fig.~1(c) for trapping $^{87}$Rb above a silver nanoshell at room temperature with $\Omega_0= 25$ GHz (corresponding to $\sim10^8 I_{\textrm{sat}}$, where $I_\textrm{sat} \approx 1.7$ mW/cm$^2$) and $\delta = 25$ THz, we estimate  a trap depth of $\sim25$ MHz and a trapping frequency of $\sim 5$ MHz.  Both the magnetic field noise and laser detuning limit the decoherence rate to $\sim 10$ Hz and the heating rate to  $\sim 1$ Hz, meaning that the atom can be trapped for $\sim$ 1 second.

The controlled patterning of arrays of metallic nanoparticles can be done  lithographically in a top-down approach or through the controlled self-assembly of metallic nanoparticles in a bottom-up approach \cite{Nagpal09,Lindquist12,fan10,Grzelczak10}.   In any nanofabricated system one must contend with disorder; the relevant disorder in this system occurs in the particle positioning and particle formation.  
In lithographic approaches one can control the particle formation at the level of 1-2 nm
\cite{Lindquist12}.  In bottom-up, self-assembly approaches it is possible to create large regions of well ordered crystal with a finite density of point and line defects, much like a conventional solid \cite{Grzelczak10}.   
Due to the local nature of the traps the disorder in the particle positioning will not affect the trapping.  Errors in the particle formation can influence the trap by shifting the plasmon resonance and the field enhancement of each particle.   To achieve consistent traps the fractional error in the plasmon resonance should be smaller than its inverse quality factor $Q= \omega_{sp}/\kappa$, which for silver(gold) nanospheres goes up to 80(20) 
\cite{Johnson72Palik85,Hartland11}.  Currently, metallic nanoshells can be made with a fractional error in the radius of less than 5$\%$, which is comparable to the inverse of $Q$ \cite{Rycenga11}. 
\begin{center}
\begin{figure}[htbp]
\includegraphics[width=.45 \textwidth]{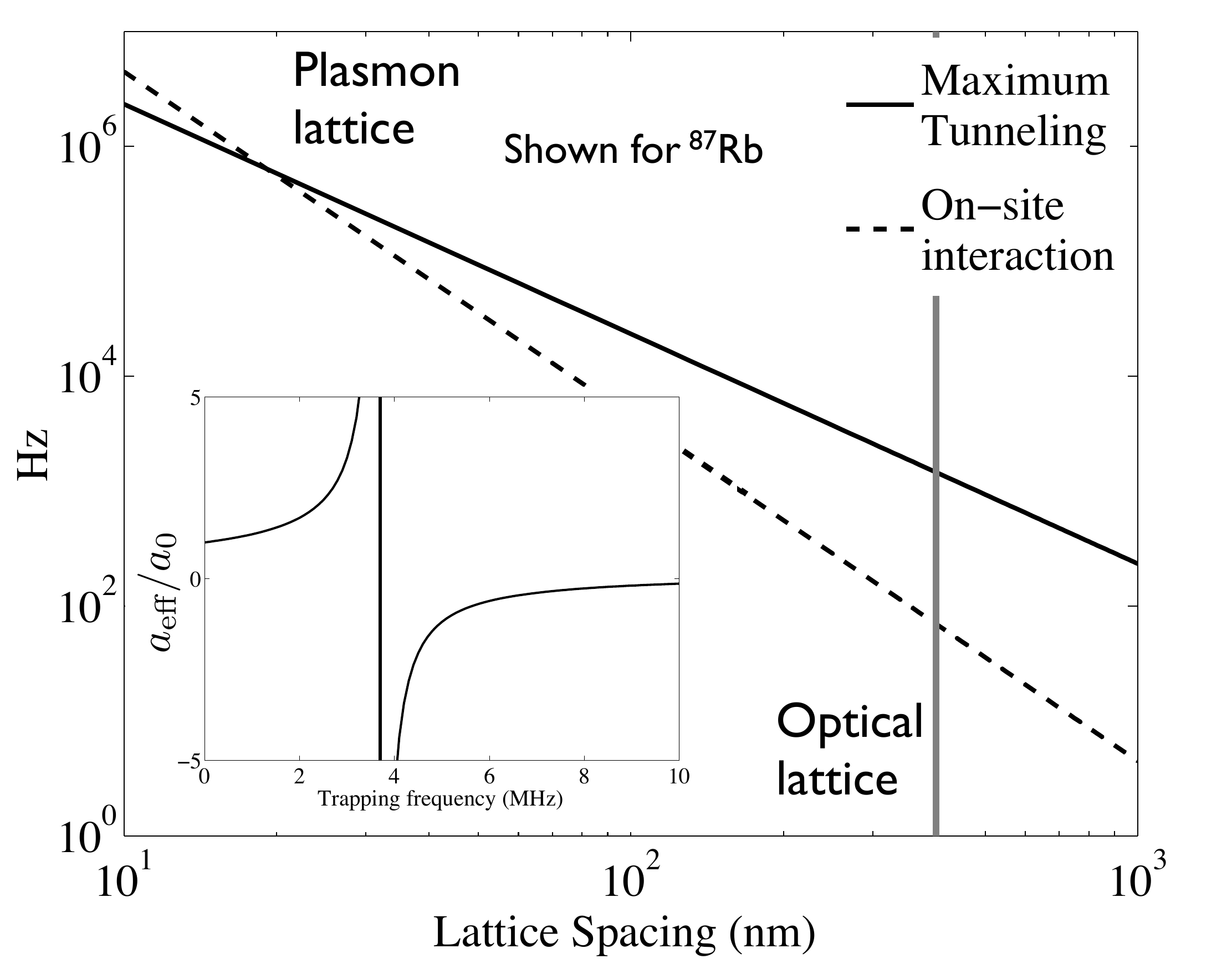}
\caption{\label{fig:U}Shows the scaling of the maximum tunneling in the lowest band, and the corresponding on-site interaction.  Calculated using the Wannier functions for a sinusoidal potential.  (Inset) Energy dependent scattering length for two $^{87}$Rb atoms on a single site as a function of the trap frequency. 
}
\end{figure}
\end{center}

As a first example application of this system we consider a realization of the single-band Hubbard model in the novel regime of large atomic density \cite{BlochDalibard08}.   
As an example, Fig.~1(d) shows that a well defined lattice potential can be achieved with a period of 60 nm,
which is within current fabrication limits. 
Figure 2(a) illustrates the scaling for the maximum tunneling in the lowest band and the corresponding on-site interaction $U_0$. 
\cite{Jaksch98}.   In the supplementary material we show that the tunneling rate can also be tuned through appropriate polarization control \cite{supp}.  

These nanoscale traps reach a regime of atomic confinement where the ground state uncertainty becomes comparable to the free space scattering length. 
  For two atoms in a  3D isotropic trap the two-body scattering problem can be solved exactly, leading to an effective scattering length $a_{\textrm{eff}}(\omega_T)$ which depends on the confinement energy  \cite{Busch98,Bolda02}.  
  The inset of Fig.\ 2 shows that a resonance emerges in the effective scattering length as a function of trap frequency \cite{supp}.  

Disorder in the lattice will also  effect  the Hubbard model.  The dominant effect arises from shifts in the local atomic potential at each sphere as the plasmonic enhancement factor changes from site to site.   From Eq.~(2) one can show that the rms of the disorder potential is given by $U_{dis} \approx \frac{\Omega^2}{2\delta} (z_T^9/a^9 Q^2) \eta/\omega_{sp}$, where $\eta$ is the rms error in the plasmon resonance.  If we take $\eta/\omega_{sp} \sim 5\%$, then for a wide range of parameters, including those in Fig. 1(d), we find that $U_{dis}$ can be made smaller than, or comparable to, the maximum tunneling.  In addition, since the disorder is static one can reduce it using the techniques described in Ref.~\cite{Pichler12}.   The effect of disorder on the single-particle physics is well understood \cite{Lagendijk09}; moreover, the interplay between interactions and disorder in the Hubbard model, as studied in Ref.~\cite{intDis1,intDis2,intDis3,intDis4}, is an interesting new regime which can be explored in the present system.

\begin{center}
\begin{figure}[thb]
\includegraphics[width=.42 \textwidth]{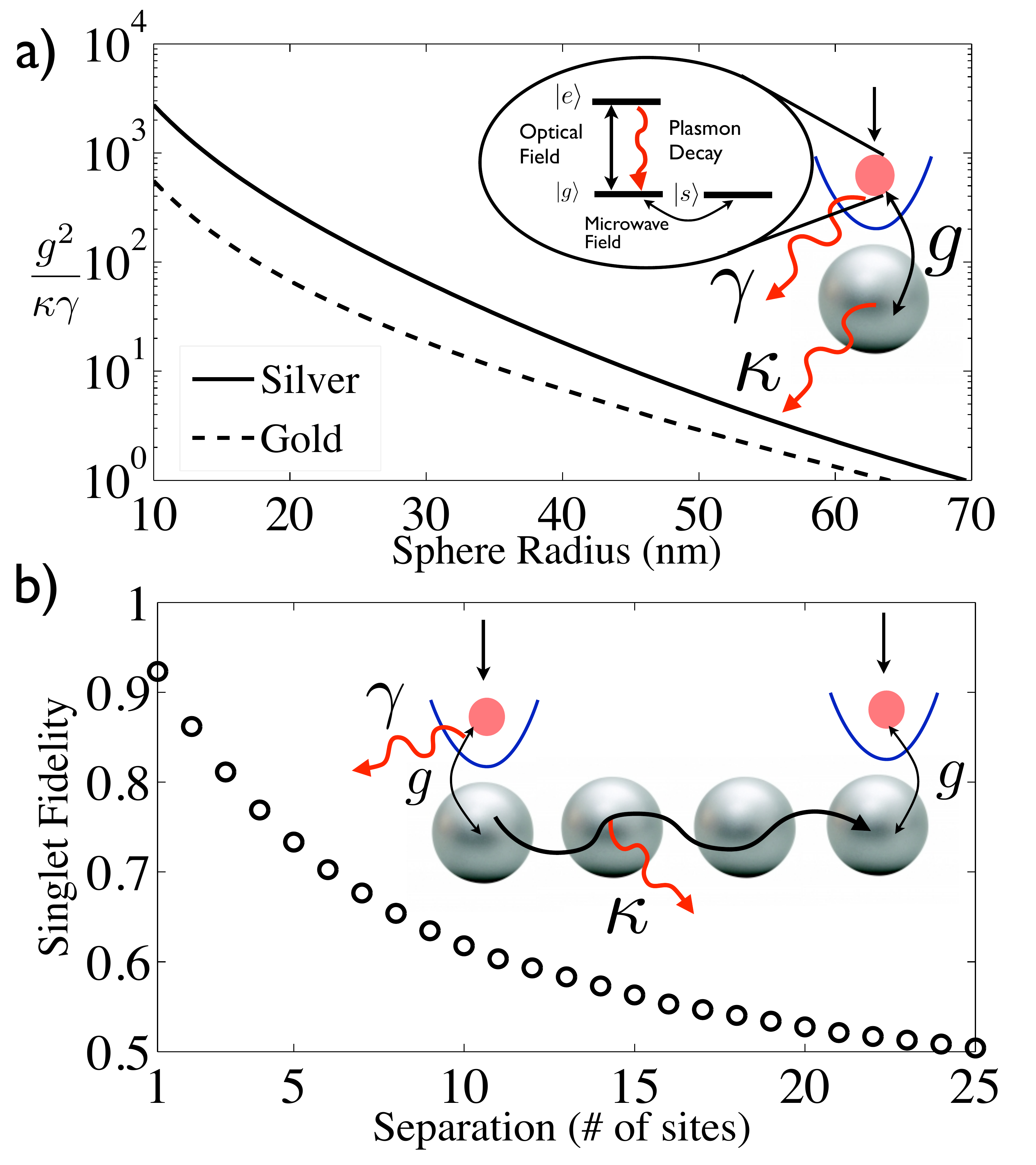}
\caption{\label{fig:RadInt}
a) Shows the cavity QED figure of merit $g^2/\kappa \gamma$ with changing system size assuming the atom is trapped at a distance of twice the sphere radius.  We show the scaling for both silver and gold nanoshells with a $Q$ of $80$ and $20$, respectively.
(Inset)  Single atom trapped above a nanosphere acts as cavity QED system with atomic and cavity losses $\gamma$ and $\kappa$, respectively, and a coherent coupling $g$.  
 b)  Fidelity for generating a ground  state singlet state between two atoms on the lattice with their separation after optimization.  The entanglement is generated through interaction with the collective plasmon modes, where we took the metal losses of bulk silver.  (Inset) Scalable cavity QED array of atoms and plasmons.
}
\end{figure}
\end{center}

We now consider long range interactions within the plasmonic lattice, associated with  the strong radiative coupling between the atoms and spheres \cite{Genov11}.   This  can be viewed as a strongly coupled cavity QED  system.
The  coupling between the atoms and the near field of the sphere is given by $g \sim \mu_0 d_0 / \epsilon_0 r^3$ where $d_0= \sqrt{\hbar \omega_{sp} \alpha(0)/2 }$ is the quantized dipole moment of the sphere  \cite{Vries98}.
Since the plasmons are overdamped the relevant coupling is given by the Purcell  factor $P=g^2/\kappa \gamma$. 
The plasmon linewidth $\kappa$ has contributions from radiative and ohmic losses.  
The radiative damping rate is $k^3 d_0^2/3 \pi \epsilon_0 \hbar \sim k^3 a^3 \omega_{sp}$.  Large spheres are radiatively broadened and, in this case, $P  \sim (k r)^{-6}$, while for small spheres  $P \sim Q a^3/k^3 r^6$.  
 In both limits, when $r \ll \lambda/2 \pi \sim 100$ nm the atoms enter the strong coupling regime $P \gg 1$, see  Fig.\ \ref{fig:RadInt}(a)  \cite{mulitpolePurcell}.  
 
For a lattice of nanospheres, intersphere coupling is also present and leads to delocalized plasmon modes in the lattice \cite{Quinten98,Krenn99}.    
We calculate the interaction of two atoms through these modes in a 1D chain of nanospheres.  For each sphere in the chain we can write the self-consistent equation for their dipole moments as \cite{multipole}
\be \label{eqn:mode}
\bm{p}_n = \alpha(\omega) \big( \bm{E}_n + \bm{N}_{nm} \bm{p}_m \big)
\ee
where $\bm{p}_n$ is the induced dipole moment of the $n$th nanopoarticle, $\bm{E}_n$ is the incident field, and $\bm{N}_{nm}$ is the 3x3 matrix that gives the dipole field at site $n$ due to the dipole at site $m$.   In 1D   two sets of transverse modes where the dipoles are oriented perpendicular to the chain and one set of longitudinal modes for parallel orientation.   
Defining $\tilde{\bm{p}}_q$  to be the $q$th eigenvector of $\bm{N}_{nm}$ with eigenvalue $D_q$, then  the effective polarizability of the $q$th mode is  $\alpha_q^{-1} = \alpha^{-1} - D_q$, i.e.
$\tilde{\bm{p}}_q = \alpha_q \tilde{\bm{E}}_q$.  For a Lorentzian polarizability the real part of $D_q$ gives the shift in the resonance frequency of the $q$th mode and the imaginary part gives the change in the linewidth.   $N_{nm}$ is diagonalized by Fourier transform and if we neglect all but nearest neighbor terms $D_q = 2 N_{01}^r \cos q - i k^3/6 \pi \epsilon_0$, where $N_{01}^r=\textrm{Re}(N_{01})$. 

Let us consider atoms trapped above the 1D array of spheres.  
The plasmonic modes can be adiabatically eliminated using standard methods in quantum optics 
\cite{Haroche82}.  For  two-level atoms polarized parallel to the 1D chain the atomic density matrix evolution is 
\begin{align} \label{eqn:master}
\dot{\rho} &= -\frac{i \omega_{at}}{2} \sum_n \comm{\sigma_n^z}{\rho}- \frac{i}{2} \sum_{n  m} \delta \omega_{nm} \comm{\sigma_n^+\sigma_m^-}{\rho} \nonumber\\
&-\frac{1}{2} \sum_{n,m} \gamma_{nm} \big( \{\sigma_n^+\sigma_m^-,\rho\}  - 2 \sigma_m^-\, \rho \, \sigma_n^+\big)  \\
\delta \omega_{nm} &= - \frac{3\,  \ell^3}{8 k^3  z^6} \Gamma_0\, \textrm{Re}\bigg( \frac{i\, e^{i q_r^* \abs{n-m}}}{ \sin q^*} \bigg)  e^{-q_i^*\abs{n-m}} \\
\gamma_{nm} &=  \frac{3\, \ell^3}{8 k^3  z^6} \Gamma_0\, \textrm{Im}\bigg( \frac{i\, e^{i q_r^* \abs{n-m}}}{ \sin q^*} \bigg)  e^{-q_i^*\abs{n-m}} 
\end{align}
where $z$ is the position of the atoms above the sphere and $q^*=q_r^*+i\,q_i^*$ is the resonant wavevector such that $\alpha_{q^*}^{-1}(\omega_a)=0$.  The first line in Eq.~(\ref{eqn:master}) describes the coherent evolution and the second line describes the collective dissipation.  Here  we have neglected the  contribution to the interaction from free-space radiative modes.

The coherent and dissipative contributions to Eq.~(\ref{eqn:master}) are equally strong when the atom and plasmon are near resonant.  Working far off resonance, however, results in  purely coherent dynamics, which can be used to implement long-range interacting spin models including frustration \cite{Strack11,Gardner10}.  Alternatively,  the collective dissipative dynamics can be used to prepare correlated atomic states \cite{Verstraete09}.
As an example, we now show how to directly prepare a ground state singlet between two atoms separated by large distances on the lattice.  We take two  ground states $\ket{g}$ and $\ket{s}$ and an excited state $\ket{e}$ which is coupled to $\ket{g}$  via an external field and  only decays via the plasmons back  to $\ket{g}$ [see inset to Fig.~3(a)].  An external microwave field mixes the two ground states.  To prepare the singlet state $\ket{S}=\ket{gs}-\ket{sg}$ we use a similar approach to Ref. \cite{Morrison08} whereby the singlet state is engineered to be the steady state of a driven, dissipative evolution.  We take a separation $n$ such that $\cos q_r^*n =1$ and
\be
\dot{\rho}=- \gamma_{0n} \mathcal{D}[\sigma_{1}^{ge}+\sigma_{2}^{ge}] \rho - \delta \gamma_n( \mathcal{D}[\sigma^{ge}_1]+\mathcal{D}[\sigma^{ge}_2] )\rho
\ee
where $\mathcal{D}[c]\rho = 1/2\{c^\dagger c,\rho\} - c \rho c^\dagger$ and $\delta \gamma_n = \gamma_{00}-\gamma_{n0}   \sim \gamma_{00}\, (\ell^3/ a^3 ) n /Q $ for $n \ll Q$.  
The dynamics can be mapped to a cavity QED system by identifying $\gamma_{0n}$ with the collective decay $g^2/\kappa$ and $\delta \gamma_n$ with the free space decay $\gamma$.  The two excited states $\ket{eg}$ and $\ket{ge}$ split into a superradiant state $\ket{eg}+\ket{ge}$ and a subradiant state $\ket{eg}-\ket{ge}$ with decay  rates $2\gamma_{0n}+\delta \gamma_n$ and $\delta \gamma_n$, respectively.  

The singlet preparation proceeds as follows.  First, we selectively excite the subradiant transition $\ket{gg}$ to $\ket{ge}-\ket{eg}$ by driving with a weak external laser field $\Omega \sim \delta \gamma_n  \ll \gamma_{00}$, which we take to have a $\pi$ phase difference on the two atoms.   Second,  in order to make the singlet state a unique steady state, we apply a global microwave field to mix the triplet ground states 
without affecting the singlet state.  
In the resulting dynamics, the pumping rate into the singlet state is $\Omega^2/\delta \gamma_n$, while the pumping rate back into the triplets is $\Omega^2/\gamma_{00}$ \cite{supp}.  The steady state of this process gives the singlet state with fidelity $F=\bra{S}\rho\ket{S}\sim1 - 1/P'$ where $P'=\gamma_{00}/\delta \gamma_n$.  Fig.~\ref{fig:RadInt}(b) shows the fidelity for two atoms with variable separation obtained from numerical simulation of Eq.~(\ref{eqn:master}).

To measure the correlations in this system,  an all optical approach could be realized by making the nanoparticle array in the near field of a solid immersion  lens (SIL), which enhances the resolution beyond the diffraction limit by a factor of $n$, the index of refraction of the SIL \cite{Wu99}.  Combining a SIL with e.g.~super resolution microscopy techniques would allow one to reach the requisite resolution of $\sim$50 nm at  optical wavelengths \cite{Huang09}.  

Our analysis shows that combining cold atom techniques with nanoscale plasmonics reaches new regimes in controlling both the collective motion of atoms and atom-photon interactions.  Combining excellent quantum control of isolated atoms with nanoscale localization, may open up exciting new possibilities for quantum control of ultracold atoms.

This work was supported by the Harvard-MIT CUA, NSF, the Physics Frontier Center, EU project AQUTE,  the ARO, DARPA OLE program, and Stanford AFOSR MURI  $\#$FA9550-12-1-0024.

\setcounter{equation}{0}
\renewcommand{\theequation}{S\arabic{equation}}
\setcounter{figure}{0}
\renewcommand{\thefigure}{S\arabic{figure}}

\newpage
\appendix
\section{Supplemental Material}

\emph{Van der Waals interaction with the Nanosphere --}
A ground state atom experiences an attractive van der Waals (vdw) force when placed near the sphere due to the virtual emission and reabsorption of photons reflected from the surface \cite{Wylie84}.  This is a purely quantum mechanical effect and can be interpreted as a modification of the Lamb shift due to the presence of the material, which changes the photon density of states.  In particular, if we write the atom-photon interaction Hamiltonian as 
\be
H_I= - \bm{\mu} \cdot \bm{E}(\bm{r}_0)
\ee
where $\bm{\mu}$ is the dipole operator and $\bm{E}$ is the electric field, then using second order perturbation theory one can write the energy shift of the ground state as 
\be
\delta E_a = -\frac{1}{\hbar} \sum_{k,e} \frac{\bra{0} E_\alpha \ket{k} \bra{k} E_\beta \ket{0} \bra{g} \mu_\alpha \ket{e}\bra{e}\mu_\beta \ket{g}}{\omega_k+ \omega_e}
\ee
where $\ket{0}$ refers to the vacuum, $\ket{k}$ to a one-photon state in the $k$th mode of the system, and $\ket{g,e}$ are the ground and excited states of the atom.  Applying Kramers-Kronig relations one can rewrite this as  \cite{Wylie84}
\begin{widetext}
\be\label{eqn:uvdw}
\delta E_a =- \frac{\hbar}{2 \pi} \textrm{Im} \int_0^\infty d\omega\, G_{\alpha \beta}(\bm{r}_0,\bm{r}_0;\omega) \alpha_{\alpha \beta}(\omega) = -\frac{\hbar}{2 \pi} \int_0^\infty d\xi\,  G_{\alpha \beta}(\bm{r}_0,\bm{r}_0;i \xi) \alpha_{\alpha \beta}(i \xi)
\ee
where we have defined the correlation functions $G_{\alpha \beta}(\bm{r},\bm{r}';t) = i \langle \comm{E_\alpha(\bm{r},t)}{E_\beta(\bm{r}',0)} \rangle \Theta(t)/\hbar$ and $\alpha_{\alpha\beta}(t)= i \bra{g} \comm{\mu_\alpha(t)}{\mu_\beta(0)} \ket{g} \Theta(t)/\hbar $.  These can be identified with the field and atomic susceptibilities, respectively.  The field susceptibility can be obtained from the classical solution for the electric field of an oscillating dipole near the sphere   \cite{Wylie84}.  The van der Waals interaction is obtained from the reflected contribution to $G_{\alpha\beta}$.  We work in the quasistatic limit where the distance between the atom and sphere is much less than a wavelength.  This results in the reflected field of a dipole $\bm{p}$ above sphere: $\bm{E}_r(\bm{r},\bm{r}';\omega)=- \nabla ( \bm{p}\cdot \nabla')\Phi_r(\bm{r},\bm{r}';\omega)$, where
\be \label{eqn:refPhi}
\Phi_r (\bm{r},\bm{r}';\omega)= - \frac{1}{4 \pi \epsilon_0} \sum_n \frac{\varepsilon(\omega)-1}{\varepsilon(\omega)+1+1/n} \, \frac{a^{2n+1}}{r'^{n+1}\, r^{n+1}} P_n\big(\cos(\theta-\theta')\big)
\ee
$\varepsilon$ is the dielectric constant of the sphere, $a$ is the radius, $\bm{r}'$ is the position of the dipole, and $P_n$ is the $n$th order Legendre polynomial.  The reflected greens function is defined  by the relation 
\begin{align}
E_r^\alpha(\bm{r},\bm{r}';\omega)&=G^r_{\alpha\beta}(\bm{r},\bm{r}';\omega) \cdot  p_\beta \\
G_{\alpha\beta}(\bm{r},\bm{r}';\omega)&= -\nabla_\alpha \nabla'_\beta \Phi_r(\bm{r},\bm{r}';\omega)
\end{align}

The van der Waals force for a ground state atom is dominated by the exchange of low-frequency, off-resonant photons.   This is to be contrasted from situation for the excited states, where the atom can emit and reabsorb real photons at the resonance frequency leading to an additional correction to the van der Waals force \cite{Chance75}.   Because of this we are justified in taking $\varepsilon \to - \infty$ in Eq.~(\ref{eqn:refPhi}), which allows us to write
\begin{align}
G_{zz}^r(\bm{r},\bm{r})&= \frac{1}{4 \pi \epsilon_0} \frac{a^3}{r^6} \frac{4-3 a^2/r^2+a^4/r^4}{(1-a^2/r^2)^3},\\
G_{xx}^r(\bm{r},\bm{r})&=G_{yy}^r(\bm{r},\bm{r})= \frac{1}{4 \pi \epsilon_0} \frac{a^3}{r^6} \frac{1}{(1-a^2/r^2)^3},\\
U_\textrm{vdW}&= - \frac{C_3}{r^6} \frac{2 a^3 \big(6 - 3 (a/r)^2+(a/r)^4\big)}{\big(1- (a/r)^2\big)^3}
= - \frac{\hbar\, \gamma}{16\, k_a^3 a^3}\frac{2 a^6 \big(6 - 3 (a/r)^2+(a/r)^4\big)}{r^6\big(1- (a/r)^2\big)^3}, \\
C_3 &= \frac{\hbar}{16 \pi^2 \epsilon_0}  \int_0^\infty d \xi \, \alpha(i \xi) = \frac{ \mean{\mu^2}}{12}.
\end{align}

\begin{figure*}[t]
\begin{center}
\includegraphics[width= .95\textwidth]{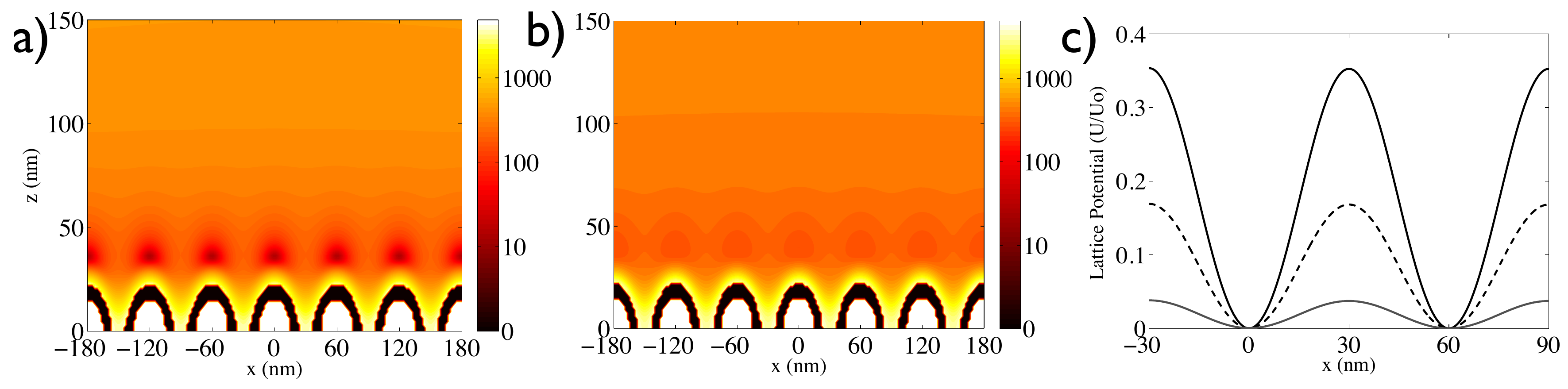}
\caption{(a-b)  Contours of atomic potential in MHz for a 1D chain of silver nanoshells including vdw with light blue-detuned to the plasmon resonance, linear polarized light is applied from the side and circularly polarized light is applied from above.  The lattice potential can be tuned by changing the polarization between linear and circular: $U_z/U_0=1$ in (a), while $U_z/U_0=0.75$ in (b). (c) Lattice potential along the chain for different amounts of circular polariztion: $U_z/U_0=1$ (solid), $U_z/U_0=0.75$ (dashed), and $U_z/U_0=0.5$ (grey).}
\label{default}
\end{center}
\end{figure*}

\end{widetext}
where  $\gamma$ is the spontaneous emission rate for the two-level atom in free space.
In the limit $a \ll r$, $U_\textrm{vdW} \sim 1/r^6$, as expected because the sphere  responds like a dipole.  In the opposite limit, when $(r-a) \ll a$ we reproduce the well known formula for the ground state shift of an atom above a perfectly conducting plane $U_\textrm{vdW} = C_3/(r-a)^3$.  
For Rb$^{87}$, $\lambda \sim 780$ nm and $\gamma = 6$ MHz, if we take a sphere with a 20 nm radius this gives the typical scale for $U_\textrm{vdW} \sim100$ MHz, which is quite substantial.

\emph{Heating Rate from  inelastic light scattering --}
Here we calculate the heating rate due the inelastic light scattering from the trapping laser including the interaction with the nanosphere.
Because of the tight trap confinement the change in motional state arises from events where a single phonon is added or subtracted to the system \cite{grimm99}.  Expanding the fields around the trap center gives the heating rate
\be
\Gamma_{\textrm{jump}} = \Gamma_\textrm{tot} \frac{E_R'}{\hbar \omega_{T,z}} \frac{\Omega^2}{\delta^2}, 
\ee
where $\Omega$ is the Rabi frequency of the trapping light, $\delta$ is the trapping laser detuning from the atomic resonance, $\omega_{T,z}$ is the trap frequency, $E_R'= 9 \hbar^2 /2 m\, z_T^2$ is an enhanced recoil energy due to the tight trap, $m$ is the mass of the atom, and $\Gamma_\textrm{tot}$ is the total spontaneous emission rate of the atom including both radiative emission and non-radiative emission into the surface plasmon modes of the sphere.  
The lifetime of the trap is approximately given by the time it takes for the atom to hop out of the trap due to such absorption processes 
\be
t_\ell \sim \frac{\hbar \Omega^2/\delta}{ \Gamma_\textrm{jump}\, \hbar \omega_{T,z}}
\ee

We express $\Gamma_\textrm{tot}= \Gamma_\textrm{rad}+\Gamma_\textrm{non-rad}$ in terms of both radiative and non-radiative contributions.  The radiative contribution can be found from the dipole moment induced in the sphere from the excited atom
\be
\Gamma_\textrm{rad} = \gamma \abs{\hat{\mu}+ \frac{\alpha(\omega_a)}{4 \pi \epsilon_0 z^3} (3( \hat{\mu} \cdot \hat{z})\, \hat{z} - \hat{\mu})}^2
\ee
The non-radiative emission arises from near field coupling of the atom to plasmon modes of the sphere.  It can be expressed as $\Gamma_\textrm{non-rad} \propto \textrm{Im}(\bm{p} \cdot \bm{E}_r(\bm{r}',\bm{r}'))$, where $\bm{E}_r$ is the field calculated in Eq.\ (\ref{eqn:refPhi}).  $\Gamma_\textrm{non-rad}$ contains both a resonant and non-resonant contributions from the dipole and multipole contributions, respectively 
\begin{widetext}
\be
\frac{\Gamma_\textrm{non-rad} }{ \Gamma_0}=\frac{6\, a^3 }{ k_a^3\, r^6} \frac{\textrm{Im} \big(\alpha(\omega_a) \big) }{4 \pi \epsilon_0\, a^3} + \frac{3}{2} \frac{1}{k_a^3 a^3} \textrm{Im}\bigg( \frac{\varepsilon-1}{\varepsilon+1} \bigg) \frac{ a^8\, (9- 11 (a/r)^2+ 4 (a/r)^4) }{ r^8\, \big(1-(a/r)^2 \big)^3}
\ee
\end{widetext}
For moderate distances from the sphere we see that $\Gamma_\textrm{non-rad}$ is  dominated by the emission into the resonant surface plasmon mode.  In addition, this emission can be substantially greater than the radiative emission.

\emph{Tuning the lattice potential --}
In order to control the tunneling rate in the Hubbard model, one needs control over the trapping potential  in the plane of the lattice.  This can be achieved through polarization control similarly to the loading procedure.  Figure S1 demonstrates this tuning in a lattice formed by linearly polarized light.  Here adding circularly polarized light lowers the potential in the plane of the lattice, while simultaneously maintaining the trap in the vertical direction.

\emph{Effective scattering length in tight traps --}
The scattering problem for two atoms in a three-dimensional isotropic trap interacting via a contact potential can be solved exactly.  We follow the approach as described in Refs. \cite{key-1,key-2} and define an energy dependent effective scattering length as $a_{eff}(E)=-\tan\eta_{0}(k)/k$.
We find the eigenvalues of the system by solving: 
\begin{equation}
\frac{a_{eff}(E)}{l}=f(E)
\end{equation}
where $l=\sqrt{\hbar/m\omega}$ is the harmonic oscillator groundstate
length and the so called 'intercept' function $f(E)$ is defined as:

\[
f(E)=\frac{1}{2}\tan\left(\frac{\pi E}{2\hbar\omega}+\frac{\pi}{4}\right)\frac{\Gamma(\frac{E}{2\hbar\omega}+\frac{1}{4})}{\Gamma(\frac{E}{2\hbar\omega}+\frac{3}{4})}
\]

\begin{center}
\begin{figure}[t]
\includegraphics[width=0.45 \textwidth]{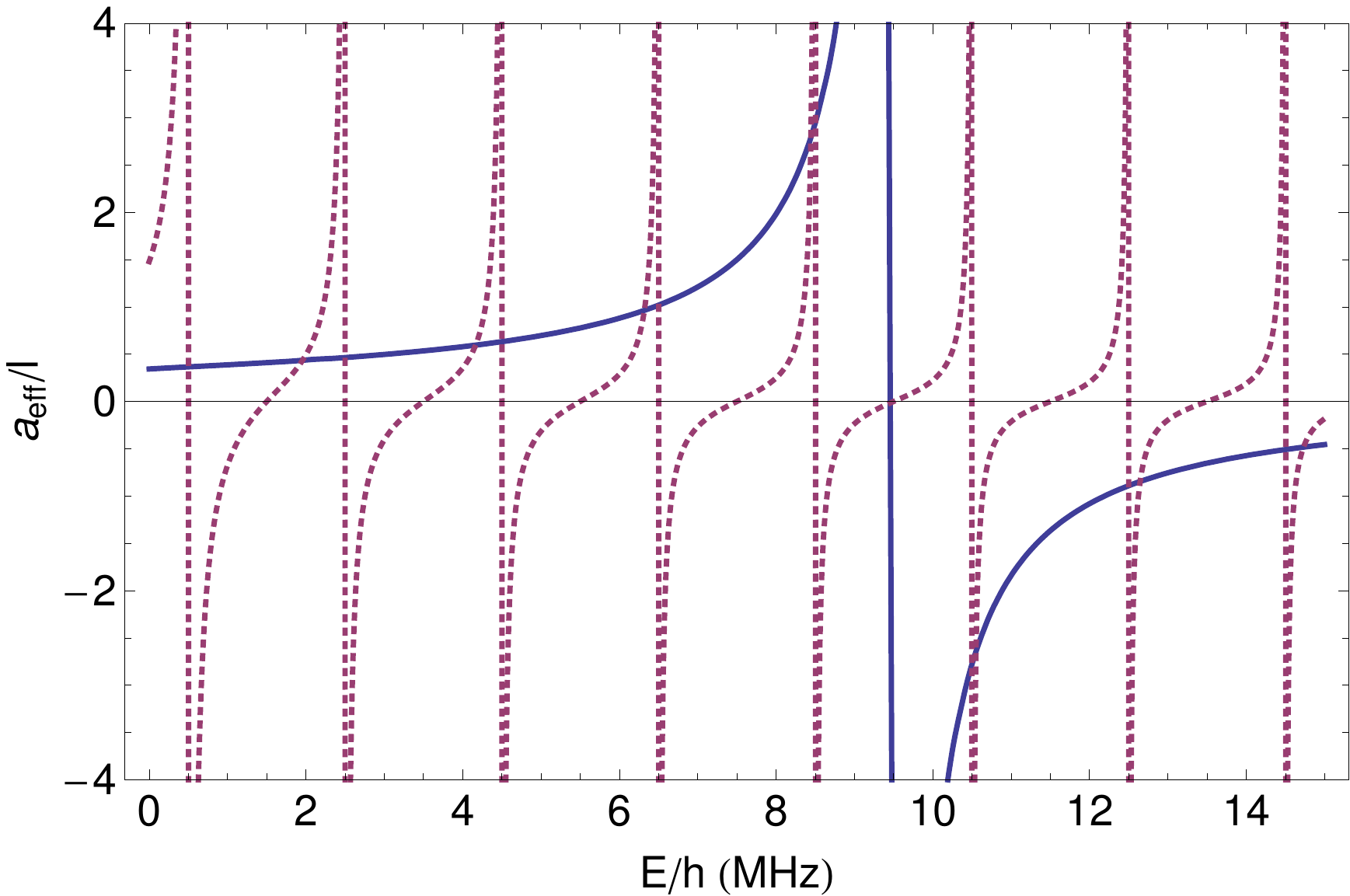}
\caption{The effective
scattering length (blue solid curve) and the intercept function (red
dashed curve) for a trap frequency of $\omega=1\,\mathrm{MHz}$.  The eigenvalues of this system correspond to the crossings of the
two curves.}
\end{figure}
\end{center}

We calculate the effective scattering length by using the accumulated
phase method as described in Ref. \cite{key-4}; we solve the radial
Schrodinger equation between $r=a_{in}=20\mathrm{\, a_{0}}$ and $r\rightarrow\infty$
where we apply the known scattering length as a boundary condition
at $r\rightarrow\infty$, this gives us the phase of the wavefunction
at $r=a_{in}$. Subsequently we calculate the effective scattering
length as a function of energy $E$ by using the phase at $r=a_{in}$
as the boundary condition.  We assume the accumulated phase is energy
independent over the energy range we consider.   This results in an energy dependent scattering length.
We verified the validity of the accumulated phase method by comparing
to the results for $^{23}$Na obtained by Bolda, et.~al. and find good agreement \cite{key-2}.
The approach breaks down if the harmonic oscillator length becomes
smaller than the van der Waals range ($l<r_{vdW}$) which is defined
as $r_{vdW}=\frac{1}{2}\left(2\mu C_{6}/\hbar^{2}\right)^{1/4}$.
For $^{87}$Rb this implies the trapping frequency should be less than $12$ MHz.

Figure S2 shows  the results of this calculation for $^{87}$Rb with a 1 MHz trapping frequency.   We took a triplet scattering length of $a_{T}=98.99a_{0}$
and $C_{6}=4698a_{0}$ \cite{key-5}.  For these parameters we find  a resonance in the effective scattering length near  $E\simeq h\times9.5\,\mathrm{MHz\simeq k_{B}\times}450\,\mathrm{\mu K}$, which is between the 4th and the 5th vibrational state.
In the inset to Fig\. 2 of the main text, we show the effective scattering length for the lowest vibrational
level as a function of trap frequency where we see a resonance at  $\omega \simeq 3.8$ MHz.

This scattering problem will be also affected by the sphere because it modifies the vdw interaction between the atoms.  However, the spheres contribution will be small compared the bare vdw, provided the typical distance between the atoms on a single site is much less than their distance to the sphere.

\begin{center}
\begin{figure*}[thbp]
\includegraphics[width= .9\textwidth]{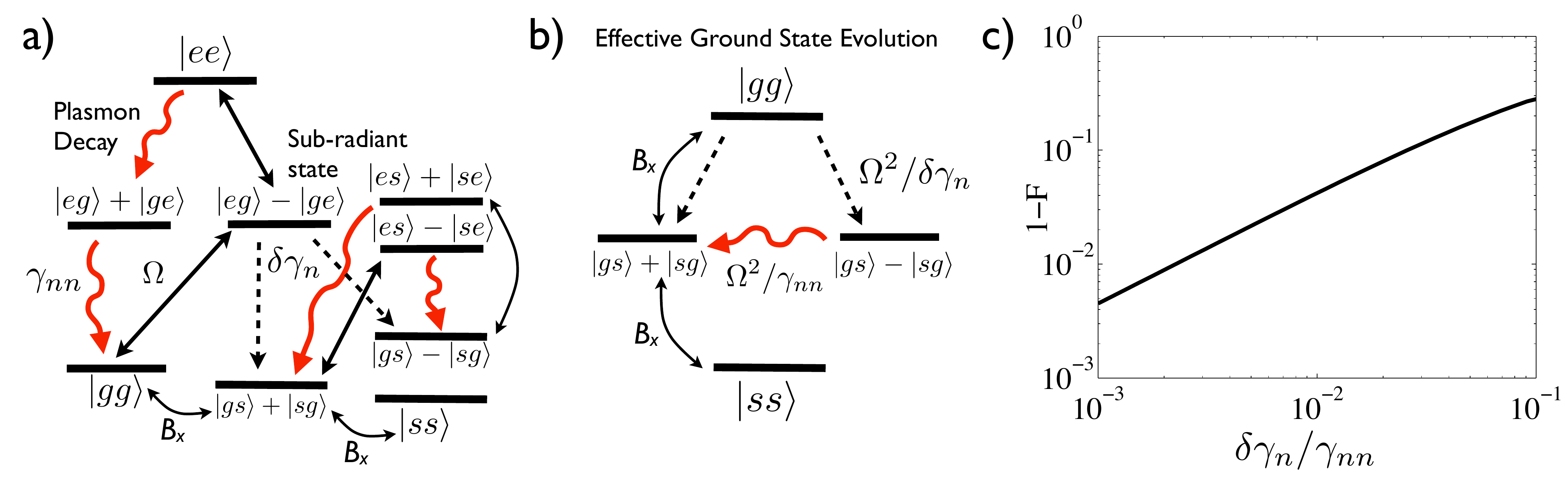}
\caption{ (a) Level diagram for two atoms showing transitions driven by external fields and decay pathways.  (b) Level diagram showing effective transition rates in the ground state manifold.  The pumping rate into the state $\ket{sg}-\ket{gs}$ is much larger than the rate out of it.  (c) Shows the infidelity for preparing the singlet state after optimizing $B_x$ and $\Omega$ as the sub-radiant states linewidth becomes narrower.  
%
}
\end{figure*}
\end{center}
\emph{Two atom Entanglement on the lattice --}
For two atoms on sites $0$ and $n$ we take the density matrix evolution
\begin{align} \nonumber
H&= \Delta( \sigma_{ee}^1+\sigma_{ee}^2) + \Omega(\sigma_{eg}^1-\sigma_{eg}^2 + h.c.) \\ 
&+B_x (\sigma_{gs}^1+\sigma_{gs}^2+h.c.) \\  \nonumber
\dot{\rho} &= -i \comm{H}{\rho} -  \gamma_{0n} \mathcal{D}[\sigma_{ge}^1+ \sigma_{ge}^2] \rho \\
& - \delta \gamma_n\big( \mathcal{D}[\sigma_{ge}^1] + \mathcal{D}[\sigma_{ge}^2]+\mathcal{D}[\sigma_{se}^1] + \mathcal{D}[\sigma_{se}^2] \big) \rho
\end{align}
where $\Delta$ is the detuning between the control fields and the excited state, $\Omega$ is an optical control field, $B_x$ is a transverse ground state magnetic field, and $D[c] \rho =1/2 \{ c^\dagger c, \rho \} -  c \rho c^\dagger$.  In addition to the decay from $\ket{e}$ to $\ket{g}$ through the plasmons we assume there is an additional decay from $\ket{e}$ to $\ket{s}$ that occurs at the rate $\delta \gamma_{n}$.  This term is essential to remove entropy from the system to cool into the singlet state. The relevant process are shown schematically in Fig. S3(a).  

The minimal error in preparing the singlet state decreases linearly with the ratio $\delta \gamma_{n}/\gamma_{nn}$ as shown in Fig. S3(c).  This can be understood in the limit of weak driving  $\Omega \ll \delta \gamma_n \ll \gamma_{nn}$.  In this limit the excited states can be adiabatically eliminated to give the effective evolution depicted in Fig. S3(b).  Because the optical pumping rate out of a state increases inversely with the linewidth, the pumping rate into the singlet state $R_{in}\approx \Omega^2/\delta \gamma_n$ can be much larger than the pumping rate out of it $R_{out}\approx\Omega^2/\gamma_{nn}$.  If, in addition, $ B_x \gg \Omega^2/\delta \gamma_n$ the triplet states are completely mixed and all triplet states can be optically pumped into the singlet state.  The ratio $R_{out}/R_{in} \sim \delta \gamma_n/\gamma_{nn}$ then determines the relative population in the triplets to the singlet state, giving the fidelity $F \approx 1- \delta \gamma_n/\gamma_{nn}$.  

As a remark we note that Eqs.~(S16-S17) can be mapped exactly to a cavity QED model by replacing $\gamma_{0n}$ by $g^2/\kappa$ and $\delta \gamma_n$ by $\gamma$, where $g$ is the coupling of a single atom to a single cavity photon, $\kappa$ is the cavity decay rate, and $\gamma$ is  the free space decay rate.  In this case the fidelity scales as $1-1/P$ where $P=g^2/\kappa \gamma$ is the Purcell factor.  This linear scaling of the singlet fidelity with the Purcell factor agrees with the limit obtained in Ref. \cite{Morrison08} using a similar dissipative approach.  The main difference between the two schemes is that in Ref. \cite{Morrison08} the cavity resonance is assumed to be far detuned from the atomic resonance, while in the present approach the two resonances are the same.  Thus they operate in qualitatively different regimes of cavity QED.  In the off resonance case the cavity interaction shifts the excited state energies for the states $\ket{eg}\pm\ket{ge}$, while in the resonant case the cavity interaction results in different linewidths for $\ket{eg}\pm\ket{ge}$.  Clearly either phenomenon  is sufficient for ground state entanglement generation.


\begin{thebibliography}{99}

\bibitem{BlochDalibard08}
I. Bloch, J. Dalibard, and S. Nascimbene, Nat. Phys. 8, 267 (2012)

\bibitem{grimm99}
R. Grimm, M. WeidemŸller, Y. B. Ovchinnikov, Adv. Atom., Mol. and Opt. Phys. Vol. 42, 95 (2000)

\bibitem{hecht}
E. Hecht, Optics, 3rd Ed., (Addison-Wesley, 1998)

\bibitem{buluta09}
I. Buluta and F. Nori, Science 326, 108 (2009)
\bibitem{jaksch05}
W. Yi, A. J. Daley, G. Pupillo, and P. Zoller, New Journal of Physics 10, 073015 (2008);
A. Gorshkov, L. Jiang, M.  Greiner, P.  Zoller, and M. D.  Lukin, Phys. Rev. Lett. 100, 093005 (2008).
\bibitem{Leung12}
V.Y.F. Leung, A. Tauschinsky, N.J. van Druten, and R.J.C. Spreeuw, arXiv 1104.3067 (2012).
\bibitem{lewenstein07}
M. Lewenstein, A. Sanpera, and V. Ahufinger, \textit{Ultracold Atoms in Optical Lattices: Simulating Quantum Many- Body Systems} (Oxford University Press, 2012)

\bibitem{Jaksch98}
D. Jaksch, C. Bruder, J. I. Cirac, C. W. Gardiner, and P. Zoller, Phys. Rev. Lett. 81, 3108 (1998).

\bibitem{Stehle11}
C. Stehle, H. Bender, C. Zimmermann, D. Kern, M. Fleischer, and S. Slama, Nat. Phot. 5, 494 (2011).
\bibitem{Murphy09}
 B. Murphy and L. V. Hau, Phys. Rev. Lett. 102, 033003 (2009);
 \bibitem{Chang09}
D. E. Chang, J. D. Thompson, H. Park, V. Vuleti\'c, A. S. Zibrov, P. Zoller, and M. D. Lukin , Phys. Rev. Lett. 103, 123004 (2009).


\bibitem{deleon12}
N. P. de Leon, M. D. Lukin, and H. Park, IEEE J. Sel. Topics Quantum Electronics 18, 1781 (2012).

\bibitem{scalableCavityQED}
J. I. Cirac, P. Zoller, H. J. Kimble, and H. Mabuchi, Phys. Rev. Lett. 78, 3221 (1997).
\bibitem{Kimble08}
H.J. Kimble, Nature 453, 1023 (2008).



\bibitem{Jackson}
J.D. Jackson, \textit{Classical Electrodynamics} 3rd Edition, Ch. 10 (John Wiley \& Sons, New York, NY, 1999).


\bibitem{Johnson72Palik85}
P. B. Johnson and R. W. Christy, Phys. Rev. B 6, 4370 (1972);
V. P. Drachev, et. al., Opt. Express 16, 1186 (2008).

\bibitem{Bohren83}
C. F. Bohren and D. R. Huffman, \textit{Absorption and Scattering of Light by Small Particles}, Ch. 5 (John Wiley \& Sons, New York, NY, 1983)

 



\bibitem{Henkel99}
C. Henkel, S. Potting, and M. Wilkens, Appl. Phys. B 69, 379 (1999).

\bibitem{supp}
See Supplemental Material for additional information on the van der Waals potential, inelastic heating rate, tuning the Hubbard model parameters, and the preparation of the singlet state through controlled dissipation.
\bibitem{Johnson11}
M. T. H. Reid, A. W. Rodriguez, J. White, and S. G. Johnson, Phys. Rev. Lett. 103, 040401 (2009).

\bibitem{Nagpal09}
P. Nagpal, N.  C. Lindquist, S.-H. Oh, and D. J. Norris, Science 325, 594 (2009)
\bibitem{Lindquist12}
N. C. Lindquist, P. Nagpal, K. M. McPeak, D. J. Norris, and S.-H. Oh, Rep. Prog. Phys. 75, 036501 (2012).



\bibitem{fan10}
J. A. Fan, C. Wu, K. Bao, J.  Bao, R. Bardhan, N. J. Halas, V N. Manoharan, P. Nordlander, G. Shvets, and F. Capasso, Science 328, 1135 (2010)
\bibitem{Grzelczak10}
M. Grzelczak, J. Vermant, E. M. Furst, and L. M. Liz-Marzan, ACS Nano 4, 3591 (2010).

\bibitem{Hartland11}
G. V. Hartland, Chem. Rev. 111, 3858 (2011).

\bibitem{Rycenga11}
M. Rycenga, C. M. Cobley, J. Zeng, W. Li, C. H. Moran, Q. Zhang, D. Qin, and Y. Xia, Chem. Rev. 111, 3669 (2011);
M. R. Jones, K. D. Osberg, R. J. Macfarlane, M. R. Langille, and C. A. Mirkin, Chem. Rev. 111, 3736 (2011).


 \bibitem{Busch98}
T. Busch, B.G. Englert, K. Rzazewski, and M. Wilkens, Foundations of Physics Volume 28, Number 4, 549-559 (1998).
 \bibitem{Bolda02}
 E. L. Bolda, E. Tiesinga, and P. S. Julienne, Phys. Rev. A 66, 013403 (2002).

\bibitem{Pichler12}
H. Pichler, J. Schachenmayer, J. Simon, P. Zoller, and A. J. Daley, arXiv 1205.6189 (2012).


\bibitem{Lagendijk09}
A. Lagendijk, B. A. Van Tiggelen, and D. Wiersma, Phys. Today 62, 24 (2009).

\bibitem{intDis1}
D. Belitz, and T. R. Kirkpatrick, Rev. Mod. Phys. 66, 261 (1994).
\bibitem{intDis2}
D. Basko, I. Aleiner, and B. Altshuler, Ann. Phys. N.Y. 321, 1126 (2006).
\bibitem{intDis3}
K. Byczuk, W. Hofstetter, and D. Vollhardt, Phys. Rev. Lett. 94, 056404 (2005).
\bibitem{intDis4}
L. Fallani, J. E. Lye, V. Guarrera, C. Fort, and M. Inguscio, Phys. Rev. Lett. 98, 130404 (2007).




\bibitem{Genov11}
D. A. Genov, R. F. Oulton, G. Bartal, and X. Zhang, Phys. Rev. B 83, 245312 (2011).

\bibitem{Vries98}
P. de Vries, D. V. van Coevorden, and A. Lagendijk, Rev. Mod. Phys. 70, 447 (1998).

\bibitem{mulitpolePurcell}
There are also multipolar corrections to the Purcell factor, but in the supplementary material we show these scale as $\textrm{Im}((\epsilon-1)/(\epsilon+1)) a^5/r^5 \sim 10^{-4}$ for silver.

\bibitem{Quinten98}
M. Quinten, A. Leitner, J.R. Krenn, and F.R. Ausenegg, Opt. Lett., 23, 1331 (1998).
\bibitem{Krenn99}
J. R. Krenn, A. Dereux, J. C. Weeber, E. Bourillot, Y. Lacroute, J. P. Goudonnet, G. Schider, W. Gotschy, A. Leitner, F. R. Aussenegg, and C. Girard, Phys. Rev. Lett. 82, 2590 (1999).

\bibitem{multipole}
Approximating the spheres as dipoles provides an accurate description of the collective plasmonic modes of nanosphere chains provided $\ell>3\,a$: S. Y. Park and D. Stroud, Phys. Rev. B 69, 125418 (2004). 

\bibitem{Haroche82}
See for example, M. Gross and S. Haroche, Phys. Rep. 93, 301 (1982).


\bibitem{Strack11}
 S. Gopalakrishnan, B. L. Lev, and P. M. Goldbart, Nat. Phys. 5, 845 (2009); P. Strack and S. Sachdev, Phys. Rev. Lett. 107, 277202 (2011).
 \bibitem{Gardner10}
 J. S. Gardner, M. J. P. Gingras, and J. E. Greedan, Rev. Mod. Phys. 82, 53 (2010).



\bibitem{Verstraete09}
F. Verstraete, M. M. Wolf, and J. I. Cirac, Nature Phys. 5, 633 (2009);
S. Diehl, A. Micheli, A. Kantian, B. Kraus, H. P. BŸchler, and P. Zoller, Nature Phys. 4, 878 (2008).  


%

\bibitem{Morrison08}
M. J. Kastoryano, F. Reiter, and A. S. S{\o}rensen, Phys. Rev. Lett. 106, 090502 (2011);

\bibitem{Wu99}
Q. Wu, G. D. Feke,  R. D. Grober, and L. P. Ghislain. Appl. Phys. Lett. 75, 4064 (1999).
\bibitem{Huang09}
B. Huang, M. Bates, and X. Zhuang, Annual Rev. of Biochem. 78, 993 (2009).


\bibitem{Wylie84}
 J. M. Wylie and J. E. Sipe, Phys. Rev. A 30, 1185 (1984), ibid., 32, 2030 (1985).
  \bibitem{Chance75}
 R. R. Chance, A. Prock, and R. Silbey, Phys. Rev. A 12, 1448 (1975).
 


\bibitem{key-1}T. Busch, B. G. Englert, K. Rzazewski, and M. Wilkens, Foundations of Physics  28, 549 (1998).

\bibitem{key-2} E. L. Bolda, E. Tiesinga, and P. S. Julienne, Phys. Rev. A 66, 013403 (2002).

\bibitem{key-4} B. J. Verhaar, E. G. M. van Kempen, and S. J. J. M. F. Kokkelmans, Phys. Rev. A 79, 032711 (2009).

\bibitem{key-5}E. G. M. van Kempen, S. J. J. M. F. Kokkelmans, D. J. Heinzen, and B. J. Verhaar, Phys. Rev. Lett, 88, 093201 (2002).

\bibitem{stirap}
K. Bergmann, H. Theuer, and B.W. Shore, Rev. Mod. Phys. 70, 1003 (1998).



\end{thebibliography}
\end{document}